\documentclass[aps,prl,reprint,superscriptaddress,10pt,nonbibnotes]{revtex4-1}

\usepackage[]{graphicx}
\usepackage[utf8]{inputenc}
\usepackage[T1]{fontenc}%
\usepackage{lmodern}%
\usepackage[hmargin=1.64cm,vmargin=1.8cm]{geometry}
\usepackage{amsmath, amssymb}
\usepackage{hyperref}%
\usepackage{xcolor}%
\usepackage{notes2bib}

\newcommand{\beq}{\begin{equation}\begin{aligned}}
\newcommand{\eeq}{\end{aligned}\end{equation}}
\newcommand{\mum}{$\mu$m}

\newcommand{\conbs}{Co$_{1/3}$NbS$_2$}

\newcommand{\tn}{\ensuremath{T_N}}

\definecolor{linkcol}{rgb}{0,0,0.4}
\definecolor{citecol}{rgb}{0.5,0,0}

\newcommand{\dqmp}{Department of Quantum Matter Physics, University of Geneva, 24 Quai Ernest Ansermet, CH-1211 Geneva, Switzerland}
\newcommand{\gap}{Group of Applied Physics, University of Geneva, 24 Quai Ernest Ansermet, CH-1211 Geneva, Switzerland}
\newcommand{\argone}{Materials Science Division, Argonne National Laboratory, Lemont, Illinois 60439, USA}
\newcommand{\epfl}{Laboratory of Physics of Complex Matter, \'Ecole polytechnique Fédérale de Lausanne, 1015 Lausanne, Switzerland}

\newcommand{\tennessee}{Department of Physics and Astronomy, George Mason University, Fairfax, Virginia 22030, USA}
\newcommand{\georgemason}{Quantum Materials Center, George Mason University, Fairfax, Virginia 22030, USA.}
\newcommand{\paulscherer}{Laboratory for Neutron Scattering and Imaging, Paul Scherrer Institut, Villigen PSI, Switzerland}
\newcommand{\maryland}{Condensed Matter Theory Center and Joint Quantum Institute, Department of Physics, University of Maryland, College Park, Maryland 20742, USA}

\hypersetup{
bookmarksopen=true,
pdftitle="Giant anomalous Hall effect in quasi-two-dimensional layered antiferromagnet \conbs",
pdfauthor="G. Tenasini et al.",
pdfsubject="Giant anomalous Hall effect in quasi-two-dimensional layered antiferromagnet \conbs", %subject of the document
pdfstartview={FitH},    % fits the width of the page to the window
pdfmenubar=true, %menubar shown
pdfhighlight=/O, %effect of clicking on a link
colorlinks=true, %couleurs sur les liens hypertextes
pdfpagemode=UseNone, %aucun mode de page
pdfpagelayout=SinglePage, %ouverture en simple page
pdffitwindow=true, %pages ouvertes entierement dans toute la fenetre
linkcolor=linkcol, %couleur des liens hypertextes internes
citecolor=linkcol, %couleur des liens pour les citations
urlcolor=blue
}

\begin{document}

\title{\texorpdfstring{Giant anomalous Hall effect in quasi-two-dimensional
%		\\
layered antiferromagnet \conbs{}}{ Giant anomalous Hall effect in quasi-two-dimensional layered antiferromagnet \conbs{}
}}

\begin{abstract}
	The  discovery of the anomalous Hall effect (AHE) in bulk metallic antiferromagnets (AFMs) motivates the search of the same phenomenon in two-dimensional (2D) systems, where a quantized anomalous Hall conductance can in principle be observed.  Here, we present experiments on micro-fabricated devices based on \conbs, a layered AFM that was recently found to exhibit AHE in bulk crystals below the Néel temperature \tn~=~29 K.  Transport measurements reveal a pronounced resistivity anisotropy, indicating that upon lowering temperature the electronic coupling between individual atomic layers is increasingly suppressed.
	The experiments also show an extremely large anomalous Hall conductivity of approximately 400 S/cm, more than one order of magnitude larger than in the bulk, which demonstrates the importance of studying the AHE in small exfoliated crystals, less affected by crystalline defects.  Interestingly, the corresponding anomalous Hall conductance, when normalized to the number of contributing atomic planes, is  $\sim \, 0.6 \; e^2/h$ per layer, approaching the value expected for the quantized anomalous Hall effect. The observed strong anisotropy of transport and the very large anomalous Hall conductance per layer make the properties of \conbs\ compatible with the presence of partially filled topologically non-trivial 2D bands originating from the magnetic superstructure of the antiferromagnetic state. Isolating atomically thin layers of this material and controlling their charge density may therefore provide a viable route to reveal the occurrence of the quantized AHE in a 2D AFM.
\end{abstract}

\author{Giulia Tenasini}
\affiliation{\dqmp}
\affiliation{\gap}
\author{Edoardo Martino}
\affiliation{\epfl}
\author{Nicolas Ubrig}
\affiliation{\dqmp}
\affiliation{\gap}
\author{Nirmal J. Ghimire}
\affiliation{\tennessee}
\affiliation{\georgemason}
\affiliation{\argone}
\author{Helmuth Berger}
\affiliation{\epfl}
\author{Oksana Zaharko}
\affiliation{\paulscherer}
\author{Fengcheng Wu}
\affiliation{\argone}
\affiliation{\maryland}
\author{J. F. Mitchell}
\affiliation{\argone}
\author{Ivar Martin}
\affiliation{\argone}
\author{L\'aszl\'o Forr\'o}
\affiliation{\epfl}
\author{Alberto F. Morpurgo}
\email{Alberto.morpurgo@unige.ch}
\affiliation{\dqmp}
\affiliation{\gap}

\date{\today}

%\pacs{}
%\keywords{}

\maketitle

The anomalous Hall Effect (AHE)  --i.e., the presence of a finite Hall resistance in the absence of an applied magnetic field, $H$-- has been long considered to be a phenomenon characteristic of metallic ferromagnets, originating from the interplay of a finite magnetization and spin-orbit interaction \cite{Nagaosa2010}. Only recently, it has been appreciated that the AHE  can arise more generally in systems with broken time-reversal symmetry (TRS). Based on this notion, the occurrence of AHE in antiferromagnets (AFMs) with vanishing magnetization has been predicted theoretically \cite{Shindou2001,Y.TaguchiY.OoharaH.YoshizawaN.Nagaosa2001,Martin2008, Chen2014} and confirmed experimentally on materials such as Mn$_3$Sn \cite{Nakatsuji2015,Li2019} \cite{Sung2018}, Mn$_3$Ge \cite{Kiyohara2016,Nayak2016}, Mn$_3$Ga \cite{Liu2017}   and Mn$_5$Si$_3$ \cite{Surgers2014,Surgers2016,Surgers2017}.
In all these systems, the  AHE originates from the non-vanishing integral of the  Berry curvature $\Omega_n(\textbf{q})$ over the occupied states, with the transverse conductivity given by $\sigma_{xy}=\frac{e^2}{\hbar}\int_{BZ}\frac{d^3 \textbf{q}}{(2\pi)^3} f[E_n(\textbf{q})] \Omega_n(\textbf{q})$, where $E_n(\textbf{q})$ is the energy of the $n$-th band in the antiferromagnetic Brillouin zone \cite{Xiao2010,Chen2014,Nayak2016,Suzuki2017,Manna2018,Manna2018a} ($f(E)$ is the Fermi function and the subscripts $xy$ denote the plane in which transport is measured).
\begin{figure*}[t] 
	\centering
	\includegraphics[width=0.8\textwidth]{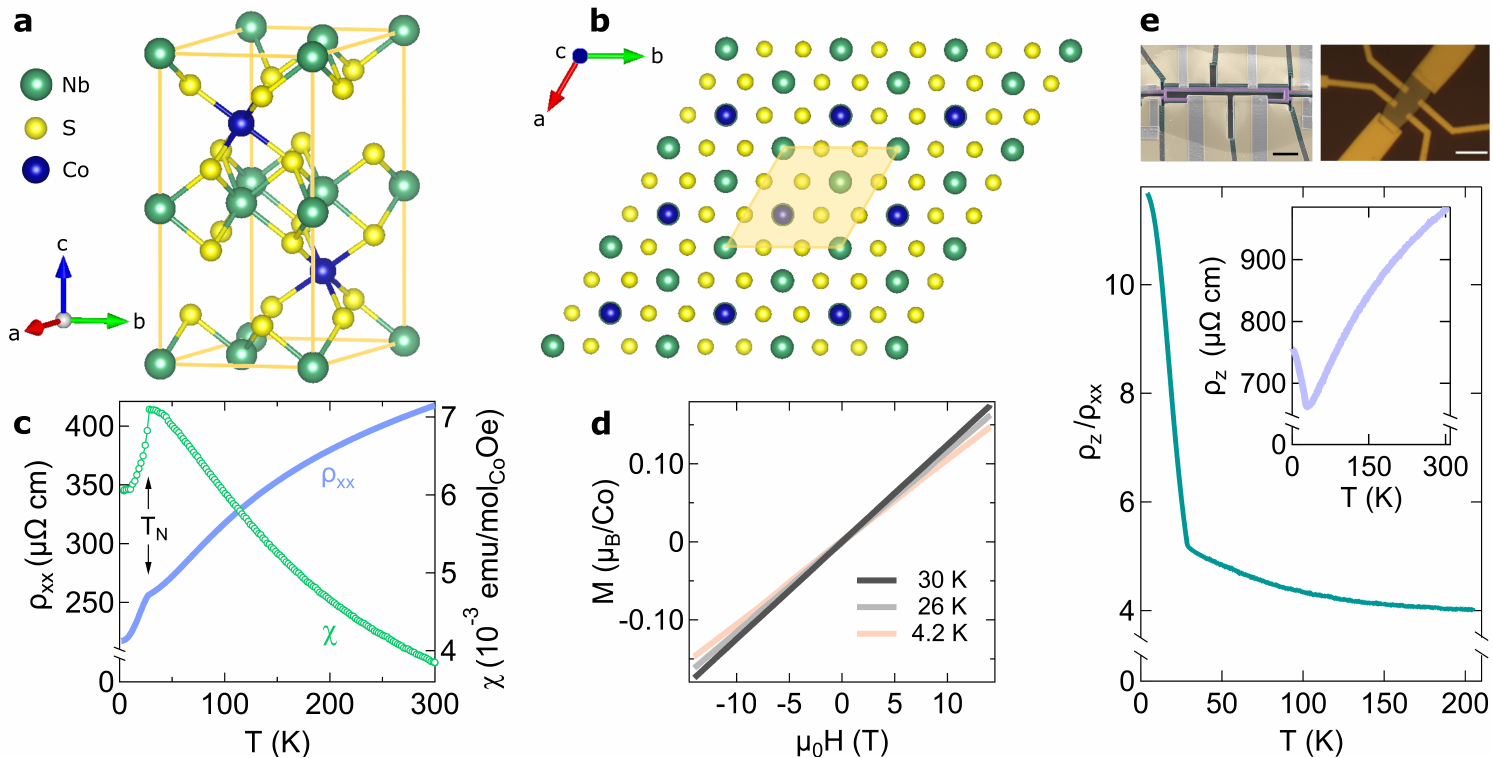}%
	\caption{\textbf{Properties of \conbs.}  \textbf{a}, Crystallographic unit cell of \conbs. \textbf{b}, Top-view of the lattice displaying the regular triangular arrangement of the Co atoms which gives rise to a $\sqrt3a_0 \times \sqrt3a_0$ superstructure  --where $a_0$ is the lattice constant of the parent compound (2H-NbS$_2$)-- typical of 1/3 fractional intercalation \cite{Nakayama2006}. The crystallographic unit cell of \conbs\ is indicated by the yellow shaded area. \textbf{c}, Temperature dependence of the in-plane resistivity, $\rho_{xx}$, (blue) and of the magnetic susceptibility, $\chi$, (green) --acquired in an applied field of 0.1~T along the $c$-axis-- both measured in a bulk crystal. The two curves show a clear kink in correspondence of the antiferromagnetic transition temperature at \tn~$\simeq$~29~K. \textbf{d},~Magnetization, $M$, as a function of magnetic field ($H \parallel c$) for different temperatures above and below \tn \  measured in  a bulk crystal; $M$ is negligible at $H = 0$. \textbf{e}, Optical micrographs of two devices investigated in the present work: a device sculpted using a  focused ion beam (top left; the scale bar is 20~$\mu$m) and a device based on an exfoliated crystal approximately 50~nm thick (top right; the scale bar is 5~$\mu$m). Temperature dependence of the ratio between the resistivity out-of-plane ($\rho_{z}$) and the resistivity in-plane ($\rho_{xx}$), showing a pronounced anisotropy below \tn. The inset shows the resistivity along the $c$-axis, $\rho_{z}$, as a function of temperature.}
	\label{fig:1}
\end{figure*}%
\\
\indent 
The possibility to express the Hall conductivity in terms of the Berry curvature underscores the topological origin of the phenomenon \cite{Nagaosa2010}. Accordingly, dimensionality is expected to play a key role in determining how the AHE manifests itself, because --as it is well-established-- three-dimensional (3D) systems with broken TRS are all topologically equivalent unless additional crystalline symmetries are imposed, whereas in two dimensions topologically distinct states can exist \cite{Yang2014}. Indeed, in two dimensions the integral of the Berry curvature of an individual band $n$ is a topological invariant (the Chern number $C_n$) that can assume any possible integer value. In two-dimensional (2D) magnetic systems, therefore, topologically non-trivial bands can occur that --whenever completely filled-- give a quantized contribution $C_n \frac{e^2}{h}$ to the measured Hall conductivity \cite{Haldane1988,Ohgushi2000,Wang2013,Chang2016,Liu2016,Tokura2019}. Such a quantum AHE has been reported in thin films of a ferromagnetic topological insulator \cite{Chang2013} (for related interesting new developments see also \cite{Serlin2019,Deng2019}), but in 2D AFMs the conditions to observe the same phenomenon have never been realized experimentally, since all known antiferromagnetic materials hosting a large AHE display rather isotropic transport properties  \cite{Nakatsuji2015,Kiyohara2016,Nayak2016,Surgers2017}. Here, we perform transport measurements on exfoliated crystals of the layered AFM \conbs\ and show that this system exhibits an anomalous Hall conductance as high as 0.6 $e^2/h$ per atomic layer as well as a strongly anisotropic transport, compatible with the existence of topologically non-trivial 2D bands.
\\
\indent 
\conbs\ is a well-known AFM, consisting of magnetic Co ions intercalated between NbS$_2$ planes, as represented in Fig.~\ref{fig:1}a.  Past transport and magnetization measurements on bulk crystals \cite{Barisic2011} (reproduced on the crystals used here, see  Fig.~\ref{fig:1}c) established that the transition to the antiferromagnetic state occurs at $T_N\simeq$~29~K (the value of $T_N$ slightly varies depending on the exact stoichiometric composition in intercalated compounds), and that for $T< T_N$ the system remains metallic, with vanishing magnetization at zero applied magnetic field (Fig.~\ref{fig:1}d).
Neutron diffraction experiments were performed \cite{Parkin1983}, and were fit to a collinear AFM structure with the magnetic unit cell double the size of the structural unit cell. In those experiments all six symmetry-related in-plane $q$ vectors were observed with nearly equal weights, and attributed to a multi-domain structure with each domain having single-$q$ ordering. However, subsequent experiments performed on this material
\bibnote[note1]{In the course of this work in order to determine the full magnetic structure of \conbs \ we have performed neutron diffraction experiments on the single crystal diffractometer Zebra at Swiss Spallation Source SINQ. We found that the data are compatible with the collinear multi-domain structure proposed by \cite{Parkin1983} but also with a single domain multi-q structure.} could not distinguish between the multi-domain single-$q$ AFM and single domain multi-$q$ AFM. As we will argue, this distinction is critical: based on symmetry arguments, the AHE is incompatible with a simple collinear AFM with single-$q$ lying in the $a$-$b$ plane that was proposed in \cite{Parkin1983}. On the other hand, it is naturally expected in a non-coplanar AFM with three-$q$, all lying in the $a$-$b$ plane. Despite being more complex than the single-$q$ state, such a multi-$q$ AFM has no uniform spin magnetization, with AHE arising due to the  large uncompensated Berry curvature in the band structure \cite{Shindou2001,Y.TaguchiY.OoharaH.YoshizawaN.Nagaosa2001,Martin2008,Kubler2014,Nakatsuji2015,MacHida2010,Ndiaye2019}. 
Indeed,  AHE  was observed recently at $T<T_N$ in bulk crystals \cite{Ghimire2018}, despite the absence of any sizable magnetization. This observation, as well as the layered nature of \conbs, motivates us to study the AHE in detail using exfoliated crystals.
\begin{figure*}[t]
	\centering
	\includegraphics[width=0.8\textwidth]{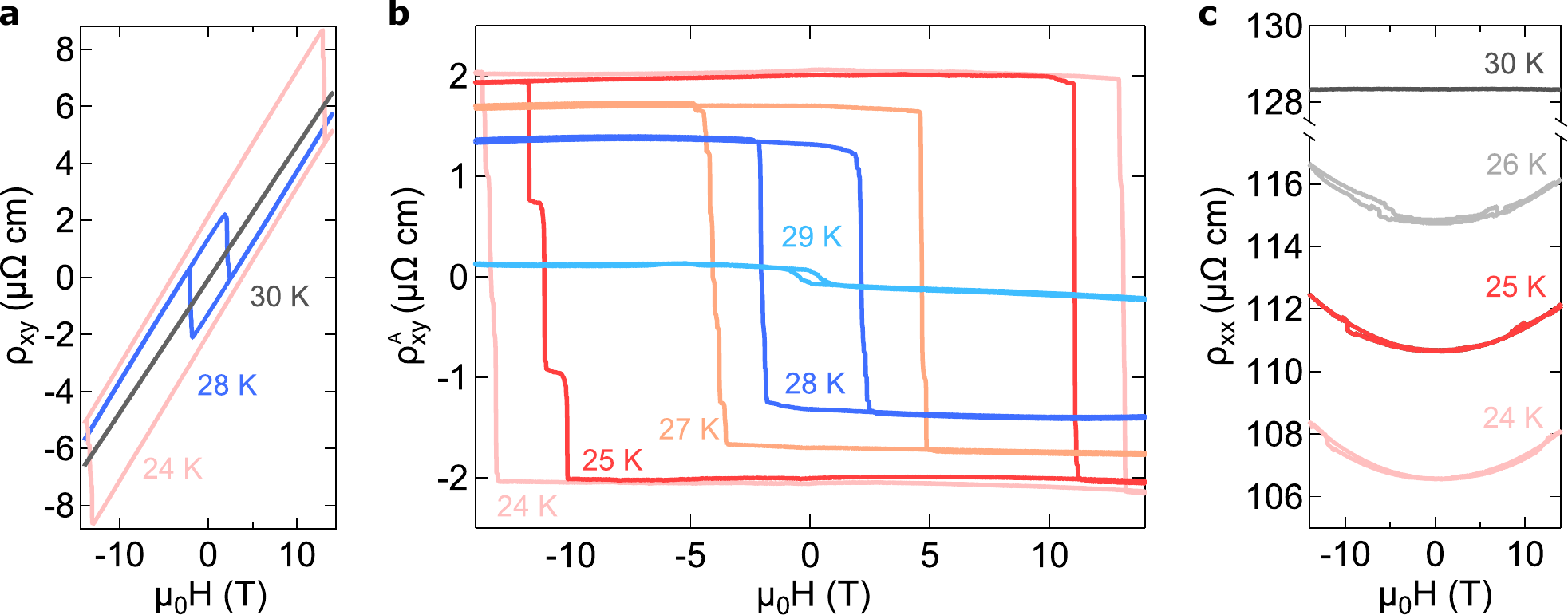}%
	\caption{\textbf{AHE in \conbs.}
		(\textbf{a}) Hall resistivity, $\rho_{xy}$ and (\textbf{b}) anomalous Hall resistivity, $\rho_{xy}^A$, measured in a representative device upon cycling the magnetic field back and forth from -14 T to +14 T ($H \parallel c$), for selected temperatures above and below \tn. The anomalous Hall resistivity in \textbf{b} is obtained by subtracting from the curves shown in \textbf{a} the linear contribution due to the ordinary Hall effect. $\rho_{xy}^A$ exhibits sharp jumps at values of coercive fields, $H_c$, that rapidly grow upon lowering $T$. \textbf{c}. Longitudinal resistivity,  $\rho_{xx}$, as function of magnetic field, acquired simultaneously to $\rho_{xy}$; a small magnetoresistance (MR~$\sim$~2\%) is observed below \tn. The small jumps visible in the curves in correspondence of $H_c$ (less than 10\% of the absolute change in $\rho_{xy}^A$) can be attributed to small material inhomogeneities mixing the longitudinal and transverse resistivity.	
	}
	\label{fig:2}
\end{figure*}%\begin{figure*}%[t]
\\
\indent 
For our transport studies we realized micro-fabricated devices based on \conbs\ crystals, either exfoliated with an adhesive tape (Fig.~\ref{fig:1}e, top right), or sculpted using a focused ion beam (FIB) \cite{Moll2018} (Fig.~\ref{fig:1}e, top left; for more details see Supplemental Material Sec.  S1). Being much smaller than bulk crystals, exfoliated layers are considerably less affected by structural defects, facilitating the observation of the intrinsic properties of \conbs. For instance, as we show below and of key importance for our results, the in-plane resistivity, $\rho_{xx}$, measured on exfoliated layers is found to be significantly lower than in the bulk \cite{Ghimire2018}, a clear indication of the reduction of scattering from structural defects. Micro-fabricated devices also allow us to compare transport along specific crystallographic directions, which is essential to reveal a pronounced anisotropy. We find that the out-of-plane resistivity, $\rho_{z}$,  (see inset in Fig.~\ref{fig:1}e), is more than one order of magnitude larger than the in-plane resistivity, with the ratio $\rho_{z}/\rho_{xx}$ increasing upon cooling, without saturating at the lowest temperature reached in the measurements ($T \simeq 4$ K; see Fig.~\ref{fig:1}e).
\\
\indent 
The AHE is readily detected by measuring the Hall resistivity,  $\rho_{xy}$, upon sweeping the magnetic field applied parallel to the $c$-axis from $\mu_0 H =-14$~T to $\mu_0 H = +14$~T and back. The data in Fig.~\ref{fig:2} are representative of the behavior observed in five devices based on exfoliated crystals having different thickness, $t$, (varying between 40 and 90~nm) and lateral dimensions (ranging from approximately $4 \times 8$~\mum$^2$ to $15 \times 26$~\mum$^2$). For temperatures $T$ above $T_N$ ($T = 30$~K in Fig.~\ref{fig:2}a), $\rho_{xy}$ exhibits a $H$-linear ordinary Hall effect (grey curve), but as $T$ is lowered below $T_N$ (blue curve measured at $T=28$~K), a pronounced hysteresis becomes visible, such that $\rho_{xy} \neq 0$ at $H = 0$. Upon lowering $T$, the coercive magnetic field, $H_c$, rapidly increases (pink curve), and below 24~K becomes larger than the highest field accessible in our experimental set-up (14~T), so that the hysteresis cycle can no longer be observed.
\\
\indent 
In this temperature range, the anomalous contribution of the Hall resistivity, $\rho_{xy}^{A}$, can be straightforwardly isolated by subtracting the linear contribution due to the ordinary Hall effect, as shown in Fig.~\ref{fig:2}b. Interestingly, despite the presence of large, sharp jumps in $\rho_{xy}^{A}$ at $H_c$, the longitudinal resistivity, $\rho_{xx}$, does not exhibit significant variations (see Fig.~\ref{fig:2}c): only a small magnetoresistance (MR~$\sim$~2\%) originating from the presence of the AFM state is observed (there is no MR at 30 K, see dark grey curve), with no significant change at $H_c$.  The comparison of the $H$-dependence of the anomalous Hall resistivity (Fig.~\ref{fig:2}b) and of the bulk magnetization (Fig.~\ref{fig:1}d) shows that $\rho_{xy}^A$ is not simply proportional to $M$, i.e., the AHE in \conbs\ is very different from that commonly observed in ferromagnets.
\begin{figure*}[t]
	\centering
	\includegraphics[width=0.75\textwidth]{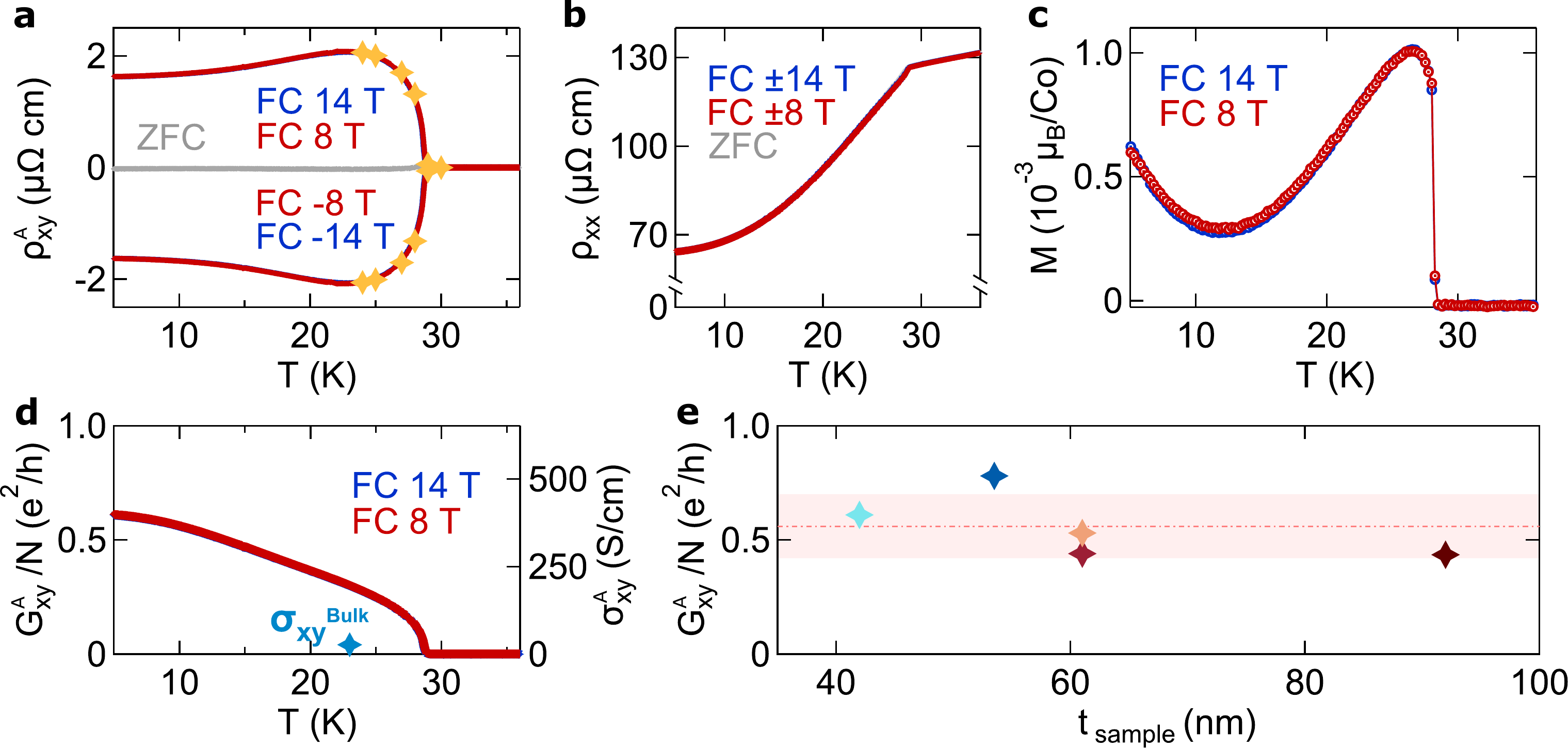}%
	\caption{\textbf{Temperature evolution of AHE in an exfoliated \conbs \ device.}
	(\textbf{a}) Anomalous Hall resistivity, $\rho_{xy}^A$,  and (\textbf{b}) longitudinal resistivity, $\rho_{xx}$, as a function of temperature measured in a representative device, after field-cooling (FC) in  $\mu_0H_{FC} = \pm$8~T (red),  $\pm$14~T (blue) and 0~T (ZFC, grey curve), with $H \parallel c$. The yellow diamonds in \textbf{a} represent the values of $\rho_{xy}^A$ extracted at  $H = 0$ from measurements of the hysteresis loops done at fixed $T$ (see Fig.~\ref{fig:2}b). $\rho_{xx}$ shows no dependence on $H_{FC}$: the curves in \textbf{b} fall exactly on top of each others and an identical $T$-dependence is observed at ZFC.
	\textbf{c}, Remnant magnetization, $M$, as a function of temperature for $H_{FC}$ of 8~T (red) and 14~T (blue)  with $H \parallel c$. \textbf{d}, Temperature dependence of the corresponding anomalous Hall conductivity, $\sigma_{xy}^A$, for FC in 8~T (red) and 14~T (blue). The blue diamond represents the maximum value of the anomalous Hall conductivity reported previously in bulk crystals (27~S/cm at 23~K, see \cite{Ghimire2018}). The right axis indicates the anomalous Hall conductance per atomic layer, $G_{xy}^A/N$, obtained normalizing the measured  $\sigma_{xy}^A$ (left axis) to the total number of contributing crystalline planes \bibnote[note2]{The anomalous Hall conductivity normalized to the number of  crystalline unit cells (each composed by two NbS$_2$ planes with intercalated Co atoms) would give an anomalous Hall conductance larger than $e^2/h$ per layer. Here, we normalize to the number of  NbS$_2$ planes since each atomic layer --if considered as a fully isolated entity-- can give rise to 2D topologically non-trivial bands in the antiferromagnetic state.}.
	\textbf{e}, Anomalous Hall conductance per atomic layer, $G_{xy}^A/N$, measured at 5~K for different samples as function of their thickness. The pink dashed line and the shaded area represent the mean value and the standard deviation of all measured values.
	}
	\label{fig:3}
\end{figure*}%
\\
\indent 
The rapid increase in coercive field upon lowering $T$ prevents the hysteresis cycle in $\rho_{xy}$ to be measured at low temperature. Therefore, to investigate systematically the evolution of the AHE,  we follow the strategy used earlier in  Mn$_3$Sn \cite{Nakatsuji2015} and Mn$_3$Ge \cite{Kiyohara2016}, and perform measurements of  $\rho_{xy}$ as a function of increasing $T$ at zero magnetic field, after cooling down the devices from $T>T_N$ to $T=5$~K in the presence of an applied field.
Irrespective of the specific antiferromagnetic state responsible for the occurrence of the AHE, which is, in general, different in different materials, the idea is as follows. If the device is cooled in the absence of an applied field, chiral microscopic domains (that we refer to hereafter as \emph{microdomains}) of the antiferromagnetic state contributing  to the AHE with different sign form with equal probability, and coexist in the material as $T$ is lowered below $T_N$. As a result, the total anomalous Hall conductivity vanishes. On the other hand, field-cooling (FC)  in a sufficiently large applied magnetic field from $T>T_N$ down to low temperature tends to align the microdomains --favoring one chirality-- making $\rho_{xy}^A$ finite. The microscopic mechanism of coupling between the antiferromagnetic microdomains and the applied magnetic field may vary. The minimal coupling that is expected to be always present is to the electronic orbital magnetization \cite{NiuOM} (along the $c$-axis) which generically accompanies the AHE, both being caused by the band Berry curvature. Another  possibility is the coupling to a weak out-of-plane spin canting due to spin-orbit interaction.  
\\
\indent 
The temperature dependence of the anomalous Hall resistivity measured after field-cooling at $\mu_0H_{FC}= \pm 8$~T (red lines) and $\mu_0H_{FC}= \pm 14$~T (blue lines) is shown in Fig.~\ref{fig:3}a, for one of our \conbs\ devices, which exhibits a representative behavior. Below $T_N$ a finite $\rho_{xy}^A$ emerges, which follows exactly the same temperature dependence for both field values. As expected for field induced alignment of antiferromagnetic microdomains, cooling in fields of opposite polarity selects microdomains of opposite chirality, resulting in equal and opposite values of $\rho_{xy}^A$. The soundness of the FC procedure is illustrated by the perfect agreement between the $T$-dependent $\rho_{xy}^A$ curve measured after field-cooling and the values of $\rho_{xy}^A$ extracted at  $H = 0$ in measurements of the hysteresis loops done at fixed $T$ (see the yellow diamonds in Fig.~\ref{fig:3}a). As expected, no AHE is observed upon cooling down the sample in the absence of an applied magnetic field (i.e., upon zero-field-cooling, ZFC, see grey curve in Fig.~\ref{fig:3}a).
\\
\indent 
We measured the magnetization of the \conbs\ bulk crystals used for exfoliation at $H = 0$ after FC (Fig.~\ref{fig:3}c), and found an extremely low residual magnetic moment smaller than 10$^{-3}$ $\mu_B/Co$ for $T<T_N$, which may originate either from a small spin polarization in the antiferromagnetic state \cite{Ghimire2018} or from the orbital magnetization of the conducting electrons \cite{Chen2018}. Irrespective of the origin of the residual moment, finding a very large anomalous Hall resistivity in the presence of a negligibly small magnetization confirms that the mechanism responsible for the AHE observed in \conbs\ is different from that at work in ferromagnets.
Note that the longitudinal resistivity  measured simultaneously to $\rho_{xy}^A$ on the same device, is entirely insensitive to FC: the exact same $T$-dependence is observed irrespective of the value and the polarity of $H$ applied during FC (see Fig.~\ref{fig:3}b; an identical $T$-dependence for $\rho_{xx}$ is observed at ZFC). Note also that the low temperature longitudinal resistivity measured on the exfoliated layers is approximately four times smaller than in the bulk crystals used for exfoliation (compare Fig.~\ref{fig:3}b and Fig.~\ref{fig:1}c).
\begin{figure*}[t]
	\centering
	\includegraphics[width=0.8\textwidth]{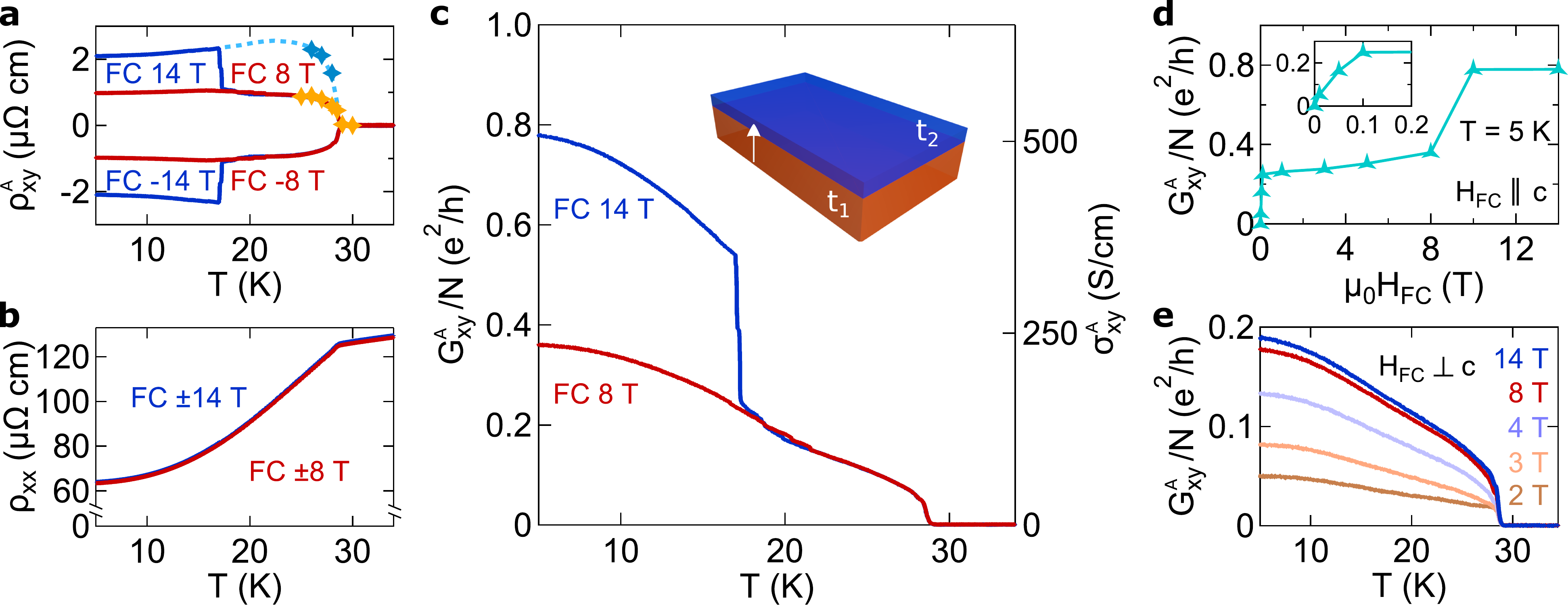}%
	\caption{\textbf{Detecting structural domain switching in the AHE.} Temperature dependence of (\textbf{a})  anomalous Hall resistivity, $\rho_{xy}^A$, and (\textbf{b}) longitudinal resistivity, $\rho_{xx}$, after FC in $\pm$8~T (red) and $\pm$14~T (blue) with $H_{FC} \parallel c$, measured in one of our devices.  In \textbf{a}, the red curves display a trend similar to the one observed in Fig.~\ref{fig:3}a, whereas the blue curves measured at  $\mu_0 H_{FC}$~= 14~T exhibit a larger value and a jump at 17~K with a drop to the value measured at $\mu_0 H_{FC}$ = 8~T. In analogy with Fig.~\ref{fig:3}a, the yellow (light-blue) diamonds correspond to the values of $\rho_{xy}^A$ extracted at  $H = 0$ (14 T) in measurements of the hysteresis loops done at fixed $T$ (see Supplemental Material Fig.~S2b). The light-blue dashed line is a guide to the eye. In \textbf{b}, $\rho_{xx}$ shows no sizable variation in correspondence of the jump observed in $\rho_{xy}^A$, indicating that at the microscopic level the magnetic state of the system remains the same across the jump.  \textbf{c}, Temperature dependence of the anomalous Hall conductance per atomic layer, $G_{xy}^A/N$, for FC in 8~T (red) and 14~T (blue). The inset shows a sketch of the envisioned magnetic configuration of the exfoliated crystal, with a structural defect that \emph{splits} the crystal in two regions (colored in red and blue) with thickness $t_1$ and $t_2$ (the arrow indicates the direction of the $c$-axis). The two regions have opposite chirality upon FC in 8~T, whereas they have the same chirality upon FC in 14~T. \textbf{d}, $G_{xy}^A/N$ as a function of $H_{FC}$ measured at 5~K with $H \parallel c$; the inset shows a zoom-in at small  $H_{FC}$.  \textbf{e}, $G_{xy}^A/N$ as a function of temperature for different $H_{FC}$ applied in the NbS$_2$ planes ($H \perp c$). Note that the value of $G_{xy}^A/N$ upon $H_{FC}$ in 8~T corresponds approximately to the value obtained for FC in 0.05~T when $H \parallel c$.
	}
	\label{fig:4}
\end{figure*}
\\
\indent 
To discuss the AHE data quantitatively we invert the resistivity tensor to calculate the anomalous Hall conductivity $\sigma_{xy}^A=\frac{\rho^A_{xy}}{{\rho_{xx}}^2+{\rho^A_{xy}}^2}$. The temperature dependence of the anomalous Hall conductivity measured (at $H = 0$) after FC at 8 T and 14 T is shown in Fig.~\ref{fig:3}d. The onset of a non-vanishing $\sigma_{xy}^A$ corresponds to the Néel temperature of \conbs, and the increase in $\sigma_{xy}^A$  upon lowering $T$ exhibits a behavior analogous to that expected for the order parameter of a second order phase transition.  Notably, due mainly to the small longitudinal resistivity of exfoliated crystals, the values of $\sigma_{xy}^A$ measured in our devices reach up to approximately 400 S/cm at low temperature, more than one order of magnitude larger than the maximum value measured in \conbs\ bulk crystals (27~S/cm; see \cite{Ghimire2018}). Such a value is comparable to --or possibly even larger than-- the highest ever reported so far for the anomalous Hall conductivity of an AFM. The corresponding anomalous Hall conductance per atomic layer is $G_{xy}^A \sim 0.6$ $e^2/h$ at $T = 5$~K \cite{note2}.
Similar values are found in all the different exfoliated devices that we measured (see Fig.~\ref{fig:3}e; device-to-device fluctuations are consistent with the uncertainty in the detailed device geometry, originating from the small size of the exfoliated crystals). 
\\
\indent 
Together with  the strongly anisotropic resistivity indicative of transport occurring  in the quasi-2D regime (see Fig.~\ref{fig:1}e), the very large observed  anomalous Hall conductance per layer suggests that  topologically non-trivial 2D bands with a non-zero Chern number $C_n$ are present in the antiferromagnetic state of \conbs. The observed metallic dependence of the in-plane resistivity on temperature implies that the Fermi level is not located in an energy gap and that, accordingly, no  quantization of the anomalous Hall conductance should be  expected. Nevertheless, even if the Fermi level is located within a topologically non-trivial band, the occupied states experience the effects of Berry curvature, explaining the presence of a large AHE. In this metallic regime, the  longitudinal conductivity is typically much larger than the transverse one (as it is the case here) and the relation $\rho_{xy}=\rho_{xx}^2\sigma_{xy}$ can be used to extract the transverse conductivity from the value of $\rho_{xy}$ and $\rho_{xx}$ (which are the actually measured quantities).  With $\sigma_{xy}=\rho_{xy}/\rho_{xx}^2$, it is clear that any extrinsic disorder increasing  $\rho_{xx}$, while leaving $\rho_{xy}$ unaffected, causes the value of $\sigma_{xy}$ extracted from the measurements to be reduced. This is indeed what appears to happen in \conbs: owing to the larger resistivity, measurements  performed on bulk materials lead to a  Hall conductance per layer of approximately   0.01~$e^2/h$ --masking the possible topologically non-trivial nature of the bands-- rather than to values of the order of $e^2/h$ that we report here 
\bibnote[note3]{In view of the envisioned topological nature of the electronic bands in the magnetic state of \conbs, one should consider the possibility that the lower resistivity measured in exfoliated layers as compared to bulk crystals originates from the presence of a surface conduction contribution. Experimentally this possibility can be ruled out, because the value of the longitudinal resistivity measured in different devices, with crystal thickness ranging  between approximately 40 and 100 nm, is the same within the experimental precision, irrespective of thickness. In contrast, a significant contribution of surface conduction would manifest itself in a smaller resistivity value for thinner exfoliated crystals.}.
\\
\indent 
An additional feature is observed in one of the devices investigated. Namely, the device shows very different temperature dependencies of the anomalous Hall resistivity upon FC for field values smaller and larger than approximately $\mu_0 H \simeq$ 8 T (see Fig.~\ref{fig:4}a).
Upon cooling down in a field smaller than (or equal to) 8 T, the anomalous Hall resistivity displays the usual qualitative behavior, whereas, when the device is field-cooled in a larger field, $\rho_{xy}^A$ has a higher low-temperature value and a downward jump at $T\simeq 17$~K is observed.
In contrast, the longitudinal resistivity measured concomitantly with $\rho_{xy}^A$ remains entirely insensitive to the conditions of FC (see Fig.~\ref{fig:4}b). The corresponding anomalous Hall conductivity obtained by inverting the resistivity tensor is shown in Fig.~\ref{fig:4}c, for FC done in a low and in a high magnetic field. Consistently with the apparent presence of two states in this device, for $T<T_N$ the evolution of $\rho_{xy}^A$ upon sweeping the magnetic field up and down also exhibits an anomalous hysteresis comprising two loops (see Supplemental Material Sec. S2).
\\
\indent 
The absence of any sizable change in $\rho_{xx}$ in correspondence of the jump in $\sigma_{xy}^A$ indicates that --at a microscopic level-- the magnetic state of the system remains the same across the jump, i.e., the jump observed in  $\sigma_{xy}^A$ is not due to a change in the magnetic ordering on the atomic scale.  As the jump in the anomalous Hall conductivity is observed only in one out of five devices, the phenomenon can be attributed to the presence in that device of two macroscopic domains (that we refer to hereafter as \emph{macrodomains}) which tend to contribute to the AHE with opposite sign. A possible origin is a structural stacking fault in the exfoliated layer which \emph{splits} the device into two separate regions, as illustrated in the inset of Fig.~\ref{fig:4}c.
The magnetic coupling across the fault energetically favors states with  opposite chirality. Thus,  if FC is performed with $\mu_0 H \leq 8$~T, the interfacial interaction dominates and the macrodomains remain anti-aligned, leading to a reduced value of AHE. For a stronger magnetic field, the energy benefit of the alignment (which goes as the product of the volume of the smaller macro-domain and the magnetic field) dominates and the chiralities of the macrodomains align, leading to a larger combined AHE.  At low temperatures, even after the strong aligning field is removed, macrodomains can remain aligned, pinned by the effective Ising anisotropy of chirality. Upon heating, the barrier to realignment can be overcome by the thermal fluctuations, causing an abrupt drop in $\sigma_{xy}^A$.
\\
\indent 
The different nature of such macrodomains, as compared to the chiral microdomains discussed earlier,  present at zero field in all the devices that we investigated, can be fully appreciated by looking at the evolution of the low-temperature anomalous Hall conductance per atomic layer as a function of field applied during FC (Fig.~\ref{fig:4}d). A very small magnetic field is sufficient  to align microdomains commonly present in all devices, since FC with $\mu_0 H = 0.05$ T already stabilizes a low-$T$ value of $G_{xy}^A/N$ corresponding to a considerable fraction of $e^2/h$ per atomic layer, as visible in the inset of Fig.~\ref{fig:4}d.
On the other hand, as shown in  Fig.~\ref{fig:4}d, the macrodomains due to the stacking fault require a much larger field to be switched.  More studies are necessary to fully understand the origin of the structural macrodomains and what determines their stability. However, in contrast to most common antiferromagnetic conductors, in AFMs that (like \conbs) exhibit AHE, magnetic field and simple transport measurements allow different types of magnetic domains and their switching to be controlled and studied experimentally.
\\
\indent 
So far we focused on the effects of the out-of-plane magnetic field (i.e., $H \parallel c$). We now address the effect of an in-plane magnetic field (i.e., parallel to the NbS$_2$ layers). The temperature dependence of the AHE is shown in Fig.~\ref{fig:4}e for different values of field applied in-plane during FC, from 2 to 14 T. The value of the anomalous Hall conductance per atomic layer, $G_{xy}^A/N$, obtained by FC in a 8 T in-plane field, is slightly less than 0.2~$e^2/h$, comparable to the anomalous Hall conductance per atomic layer induced by FC in a much smaller out-of-plane field of 0.05~T. This very large disparity in field values indicates that the effect of the in-plane field can be most likely attributed to a very small misalignment of the applied field with respect to the NbS$_2$ planes, causing a finite out-of-plane component  (an explanation certainly compatible with the precision with which our devices are mounted in the experimental set-up).
Similar behavior is observed irrespective of the orientation of the in-plane field relative to the crystallographic axis of the material, i.e., an in-plane magnetic field has no significant influence, irrespective of its direction. This is an important conclusion because it further reinforces that the alignment of chiral antiferromagnetic microdomains during the FC process is due to the coupling of the applied field to the out-of-plane magnetization \cite{NiuOM}. The latter could be due to the orbital magnetization of itinerant electrons, whose orbital electronic motion takes place predominantly in the NbS$_2$ planes, as evidenced by the quasi-2D nature of transport in \conbs\ (see Fig.~\ref{fig:1}e). Such a motion gives origin to an orbital magnetization pinned to the direction perpendicular to the NbS$_2$ planes that naturally couples to an out-of-plane magnetic field and not to an in-plane one. It can also be due to a small, spin-orbit induced canting of the Co moments.
\\
\indent 
The precise origin of the AHE in \conbs\ cannot be identified at this stage, because existing neutron diffraction experiments have not yet allowed the full determination of the magnetic structure (in part due to the small size of the available single crystals).  This is preventing reliable calculations of the electronic band structure in the antiferromagnetic state of the material. We nevertheless note that the collinear antiferromagnetic structure proposed in \cite{Parkin1983} cannot lead to AHE, since it has a symmetry that corresponds to spatial translation in the direction of the ordering vector $q$ combined with time reversal. Since spatial translation cannot affect the Hall response, and time reversal flips its sign, such symmetry immediately implies zero AHE. We also note that the scenario suggested in \cite{Smejkal2019}, based on a collinear AFM with an in-plane ferromagnetic order and an out-of-plane antiferromagnetic order is incompatible with the neutron data that show antiferromagnetic order in the plane and hence cannot be used to explain the large observed AHE.
\\
\indent 
The pronounced anisotropy of transport, the coupling of the magnetic field to the orbital magnetization, and the anomalous Hall conductance per atomic layer reaching values close to $e^2/h$ all strongly suggest that the AHE originates from the presence of 2D topologically non-trivial bands induced by the antiferromagnetic state. In the truly 2D case, models predicting the occurrence of the quantum Hall effect in antiferromagnetic states with vanishing magnetization have been known for over 10 years \cite{Shindou2001,Martin2008}. An exact quantization of the Hall conductance  occurs when the chemical potential lies within a band gap, and the magnetic structure causes some of the fully occupied bands to become topologically non-trivial. This  has been demonstrated theoretically to happen for a variety of  non-collinear and non-coplanar spin  structures \cite{Shindou2001,Martin2008,Ito2017,Zhang2017,Ndiaye2019,Li2019a}. For instance, the three-$q$ non-coplanar \emph{tetrahedral} antiferromagnetic state \cite{Martin2008} in each triangular Co layer, would be consistent with the magnitude and the direction of the antiferromagnetic ordering vectors observed experimentally, as well as with the fact that all six  symmetry-related $q$ (three $\pm q$ pairs) are observed with equal weight. Our preliminary calculations based on a realistic tight-binding model \bibnote[note4]{F. Wu, and I. Martin, Unpublished} show that this state can indeed yield large AHE with very weak orbital magnetization, consistently with our experimental observations.
\\
\indent 
As we already mentioned earlier, in the case of our \conbs\ crystals, the system is metallic and the Hall conductance (per atomic layer) cannot be precisely quantized.  Even so, it can remain  comparable to $e^2/h$, as it is expected for nearly fully filled or almost empty topological bands. In this regime, small electron density variations can cause large variations in the observed anomalous Hall conductance, particularly if the Fermi level is located close to a band edge, where the Berry curvature is typically the highest. Experimentally, it would therefore be interesting to change the carrier density in \conbs\ to see whether large changes in Hall conductance can actually be detected. It may even be possible to induce full quantization, if the density variation allows to completely fill/empty all the partially filled bands. Viable strategies to change the electron density by a significant  amount include the use of ionic liquid gated devices, if the thickness of  \conbs\  can be reduced to a few layers, or intercalation of atoms (either alkali or halogen)  in between the NbS$_2$ planes (i.e., where the Co atoms also reside). Yet another route consists in investigating other NbS$_2$-based compounds intercalated with different transition metals: Fe$_{1/3}$NbS$_2$ for instance is also an antiferromagnetic conductor \cite{Gorochov1981,Nair2019}, with a different expected band filling as compared to  \conbs, owing to the difference valence of Fe as compared to Co atoms. Although challenging experimentally, all these different strategies are worth considering because, if successful, they would allow unveiling a quantum anomalous Hall antiferromagnet, i.e., an exotic state of matter that has never been observed before.
\\
\indent 
In summary, we performed transport experiments on exfoliated and micro-fabricated \conbs\ devices that provide a new perspective on the nature of the anomalous Hall effect in the antiferromagnetic state of this material. These devices enable the observation of a much larger AHE as compared to bulk crystals --with an anomalous Hall conductance reaching values close to $e^2/h$ per atomic layer-- as well as of a large anisotropy of the longitudinal resistivity, that becomes increasingly pronounced upon lowering temperature. These findings suggest that \conbs\ represents a very promising platform to search for a topological 2D quantum anomalous Hall antiferromagnet exhibiting quantized Hall conductance at zero applied magnetic field. They also show the importance of performing experiments on small exfoliated devices to reduce extrinsic effects originating from disorder. Indeed, disorder and all sources of inhomogeneity  can be drastically reduced in exfoliated crystals, simply because their volume can be easily made smaller by a factor of approximately $10^7$-$10^8$ as compared to bulk crystals commonly used to perform transport experiments . In the case of \conbs\, the reduced disorder  results in a significantly smaller value of longitudinal resistivity that translates into a value of Hall conductance per layer reaching up to a large fraction of $e^2/h$. One may expect similar effects to be equally important in the analysis and interpretation of the measured anomalous Hall effect  in other antiferromagnetic materials that have been discussed earlier.
\medskip

%\section*{Acknowledgements}
We sincerely acknowledge Alexandre Ferreira for technical assistance,  Enrico Giannini and Bhushan Hegde for helpful support in magnetization measurements, Ignacio Gutiérrez-Lezama and Marco Gibertini for fruitful discussion and Christian R\"{u}egg for support. 
E.M. and L.F. gratefully acknowledge financial support from the Swiss National Science Foundation Grant No. 200021-175836. E.M. acknowledges Philip Moll for help and training in the FIB samples preparation, supported from the Max-Planck Society. 
Work in the Materials Science Division at Argonne National Laboratory (crystal synthesis, characterization, theoretical studies) was supported by the U.S. Department of Energy, Office of Science, Basic Energy Sciences, Materials Science and Engineering Division. F.W. is also supported by the Laboratory for Physical Sciences. A.F.M. gratefully acknowledges financial support from the Swiss National Science Foundation (Division II) and from the EU Graphene Flagship project.
%\smallskip

%\section*{Competing financial interests}
The authors declare no competing financial interests.

%\bibliography{conb3s6bib}

\begin{thebibliography}{48}
\bibitem [{note2()}]{note2}%
  \BibitemOpen
  \href@noop {} {}\bibinfo {note} {The anomalous Hall conductivity
  normalized to the number of crystalline unit cells (each composed of two
  NbS$_2$ planes with intercalated Co atoms) would give an anomalous Hall
  conductance larger than $e^2/h$ per layer. Here, we normalize to the number
  of NbS$_2$ planes since each atomic layer --if considered as a fully isolated
  entity-- can give rise to 2D topologically non-trivial bands in the
  antiferromagnetic state.}\BibitemShut {Stop}%
\bibitem [{note3()}]{note3}%
  \BibitemOpen
  \href@noop {} {}\bibinfo {note} {In view of the envisioned
  topological nature of the electronic bands in the magnetic state of \conbs ,
  one should consider the possibility that the lower resistivity measured in
  exfoliated layers as compared to bulk crystals originates from the presence
  of a surface conduction contribution. Experimentally this possibility can be
  ruled out, because the value of the longitudinal resistivity measured in
  different devices, with crystal thickness ranging between approximately 40
  and 100 nm, is the same within the experimental precision, irrespective of
  thickness. In contrast, a significant contribution of surface conduction
  would manifest itself in a smaller resistivity value for thinner exfoliated
  crystals.}\BibitemShut {Stop}%
\end{thebibliography}


\end{document}